# Over 100% Light Extraction Enhancement of Organic Light Emitting Diodes Using Flat Moire Micro-Lens Array Fabricated by Double Nanoimprint Lithography Over a Large Area

*Ji Qi, Wei Ding, Qi Zhang, Yuxuan Wang, Hao Chen, and Stephen Y. Chou\**

\*Prof. S. Y. Chou
Department of Electrical Engineering
Princeton University
Princeton, New Jersey 08544, USA
E-mail: chou@princeton.edu

J. Qi, W. Ding, Q. Zhang, Y. Wang, H. Chen
Department of Electrical Engineering,
Princeton University
Princeton, New Jersey 08544, USA

Keywords: light extraction, double-imprint, micro-lens, Moire pattern, power efficiency, organic light emitting diode, nanostructures, nanoimprint, scattering efficiency

Abstract

To improve the light extraction efficiency of organic light emitting diodes (OLEDs), we developed a novel substrate, i.e., a metamaterial based flat Moire micro-lens array formed using double nanoimprint, termed "Mlens-array", consisting of a hexagonal moiré pattern pillar array. By choosing a low refractive index dielectric material for the pillar array and a high refractive index dielectric material for the layer capped on top of the pillar array, we fabricated the Mlens-array behaving as a conventional convex optical micro-lens array. The Mlens-array was fabricated on a 4-inch wafer-size glass substrate by double-cycle compositional nanoimprint lithography (NIL) which is easy for achieving high throughput fabrication in large-scale. Applying the Mlens-array substrate in a typical green-emitting OLED, the light extraction efficiency was enhanced by over 100% (2.08-fold) compared to a control device fabricated on the conventional planar glass substrate.

**Introduction**

Generally, to extract the light from OLEDs is to find ways to reduce the total reflection at the interface between materials with different refractive indexes. One way is to utilize scattering as shown in Fig. 1a: by forming some periodic or random distributed scatterer at the glass/air interface, the light beams with large incidence angle are scattered by those structures and hence not trapped by the waveguide mode in the glass substrate[1-2]. Besides utilizing scattering, micro-lens array patterned on the glass substrate is also a well-studied way to extract light from OLEDs[3-5]. Because the surface morphology of the glass-air interface was changed, the incident angle from the substrate to the air is changed. The light rays at an incident angle larger than the critical angle of a planar glass-air interface, as shown in Fig. 1b, can be coupled out by the micro-lens array.

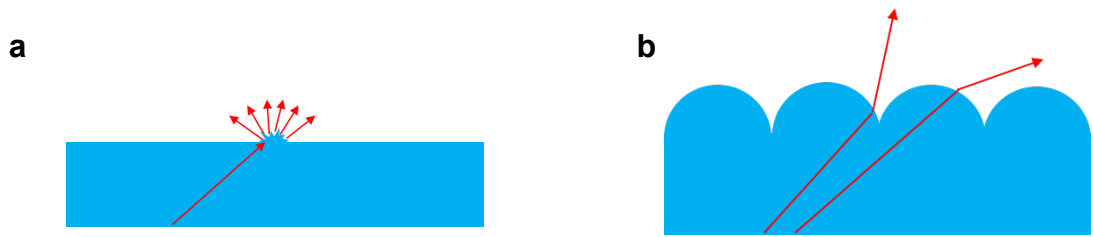

**Figure 1. Illustration of two typical light extraction mechanisms used in OLEDs.** (a) utilize scattering (b) utilize micro-lens array.

However, the fabrication of processes of the curved micro-lens are expensive and complicated, such as reflow process[6-8], laser direct-write[9-11], 3D micro-lens array[12] projection and soft lithography[13-14]. In this paper, we developed a novelly engineered metamaterial which exhibits the property as a micro-lens array. Unlike the conventional micro-lens array showing a corrugated curved surface, the metamaterial based micro-lens array has a flat structure. Hence, the fabrication of the new flat micro-lens array can be very easy and cheap. And we developed a process utilizing double-cycle compositional nanoimprint lithography (NIL) to achieve large-scale and high throughput fabrication. Because of the fabrication method we used, we called our new flat Moire micro-lens array that is formed bydouble-nanoimprint, "Moire micro-lens Array", or Mlens-array for short. By applying the Mlens-array in the OLEDs, we demonstrated the light extraction efficiency can be enhanced by over 100% compared to the control device using conventional planar glass substrate.

**Mlens-array: design and principle.** The design of Mlens-array is based on the basic working principle of optical lens and the novel properties of metamaterial in subwavelength scale. The behaviors when light rays are refracted through an optical lens can be described by the Fermat's principle[15-16], i.e., the optical length of the path followed by light between two fixed points, A and B, is an extremum, and mathematically it can be expressed as:

$$\delta S = \delta \int_A^B n \, ds = 0$$

where the optical length $S$ is defined as the physical length multiplied by the refractive index ($n$) of the material.

For example, the property of when a plan wave is focused when going through a conventional convex lens in the environment of air, as shown in Fig. 2a, can be explained by the identical optical lengths of light rays 1 and 2, i.e.,

$$nL_{11} + L_{12} = nL_{21} + L_{22}$$

For this example of a conventional optical lens, the light rays are slowed down to the same level when going through the lens, but they travel in different lengths so that they can all together reach to a same wave front at the same time. But we can also achieve the same effect in an alternative way which is the design idea of the Mlens-array: rather than making light rays travel in different lengths in the lens, we can design a lens that slows down the light rays to different levels at different positions within the lens. For example, we can design a flat dielectric plate in which the refractive index gradually decreases from the center ($n_{max}$) to the edge ($n_{min}$), as shown in Fig. 2b, and the property of focusing a plane wave can be achieved if the following relationship is satisfied,

$$n_y w + L_1 = n_{max} w + L_2$$

where $w$ is the thickness of the flat lens, $n_y$ and $n_{max}$ are local refractive indexes at the locations where light rays 1 and 2 are traveling respectively.

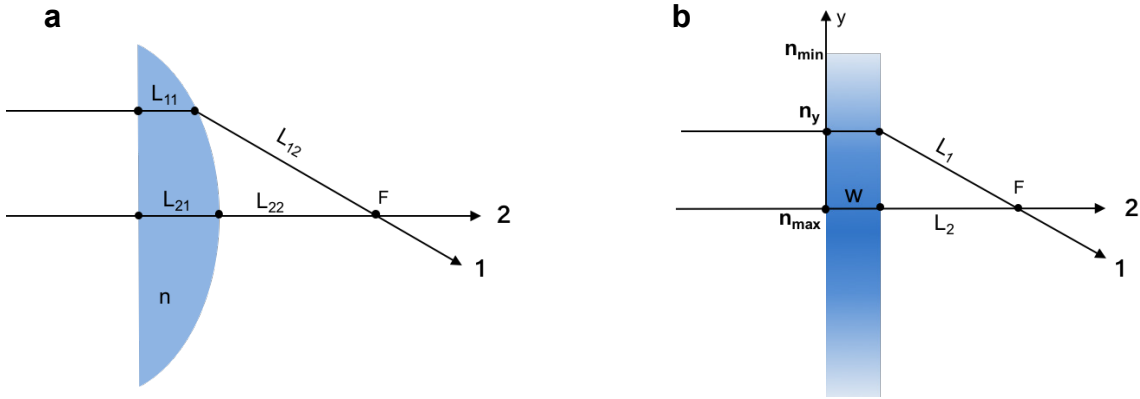

**Figure 2. Schematics illustrating two alternative lens working principles.** (a) a conventional convex lens made of homogeneous dielectric materials. (b) a novel convex lens with flat structure in which the refractive index gradually decreases from the center to the edge.

Then the next problem is how to design and fabricate such flat dielectric material with variance of refractive index. In this work, we developed a flat metamaterial micro-lens array (i.e., Mlens-array) based on the moiré pattern and the subwavelength optics. The design of Mlens-array is illustrated as follows. Suppose there are two same pillar arrays made of a dielectric material. When a pillar array pattern is overlaid on the other pillar array pattern and rotated by angle $\alpha$, a large-scale interference pattern (moiré pattern) was formed[17-18]. For example, as shown in Fig. 3a, we showed how the moiré pattern is formed from two identical hexagonal-lattice pillar arrays and it is clearly seen that the formed moiré pattern is still a hexagonal lattice. When looking at a single cell of the moiré pattern lattice, we observed that the pillars tend to overlap close to the center of cell, which results in the pillar occupation density gradually decreases from the edge to the center within in each cell. And per the mean-field theory of subwavelength optics[19-20], if the feature size of the inhomogeneities in the material is much smaller than the wavelength λ (typically < ¼ λ), the material's electromagnetic response property is an average over λ and hence the refractive index can be locally homogenized and described by a local effective refractive index <n>. Based on the discussion above, it is expected that if the moiré pattern pillar array is embedded in a flat dielectric layer with higher refractive index than that of the pillar array, within a single cell, the refractive index of the layer will gradually decrease from the center to the edge if the pillar size is much smaller than the wavelength of interest. Therefore, each cell can behave as a micro convex lens and the whole moiré pattern forms a convex micro-lens array, i.e., Mlens-array, as shown in Fig. 3b.

In an optimized Mlens-array fabricated in this work, the moiré pattern is formed by overlaying one hexagonal-lattice pillar array on another identical pillar array and the second pillar array is rotated by an angle of 2.4 degrees. The hexagonal-lattice pillar array has a lattice constant of 461nm, rhombus-shape pillars with edge length of 70nm, 300nm pillar height. Hence the periodicity p of the moiré pattern is 11um, given by

$$p = \frac{a}{\sqrt{2(1-\cos\alpha)}}$$

where a is lattice constant of the hexagonal lattice and $\alpha$ is the rotation angle. The moiré pattern pillar array is fabricated on a 1-inch square glass substrate with refractive index around 1.5 and is capped with 400nm-thick $Ta_2O_5$ with high refractive index ~2.2 to form the final Mlens-array.

a

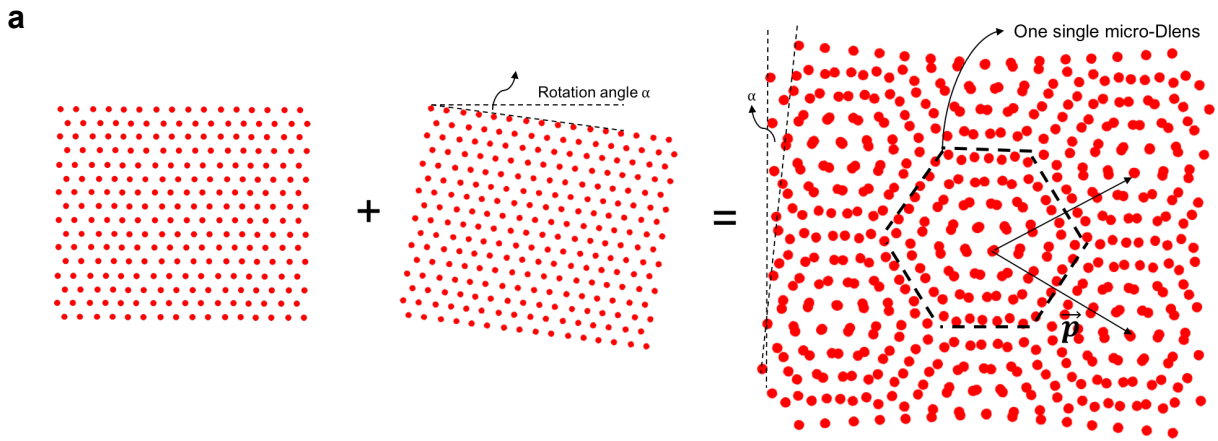

b

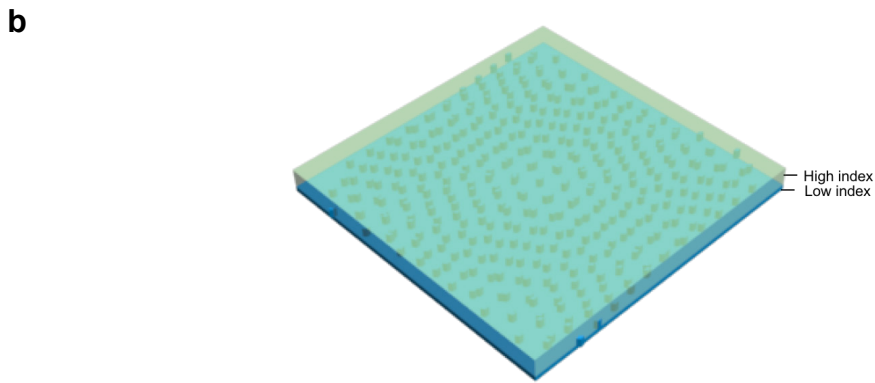

Flat convex micro-lens array

**Figure 3. Design of Mlens-array.** (a) moiré pattern is formed by overlaying one pillar array pattern on another one and the second pattern is rotated by an angle α. Each cell of the moiré pattern lattice is a micro lens. (b) a convex micro-lens array is formed if the moiré pattern pillar array is embedded in a dielectric layer with higher refractive index than that of the pillar array.

**Optical Simulations.** In order to demonstrate each single cell of the designed Mlens-array indeed behaves like a convex optical lens, the electromagnetic response of one single micro-Mlens was simulated by a commercial finite-difference time-domain solver (FDTD Solutions, Lumerical Solutions, Inc). A simulation model was constructed as shown in Fig. 4a-4c. Due to the limitation of computation power, the x-y plane feature size of the Mlens-array used in the simulation was reduced to around half of the parameters of the real design. Hence, the lattice constant of the hexagonal pillar array and the pillar diameter were chosen to be 200nm and 30nm respectively. The pillar array had a height of 300nm and a refractive index of 1.4 and was capped with 400nm-thick dielectric layer with a refractive index of 2.2. The rotation angle to form the moiré pattern is chosen to be 8 degree so that the dimension of a single micro-Mlens can be limited in a 1um square simulation zone in x-y plane to achieve a feasible simulation time. A plane wave source was placed underneath the micro-Mlens with normal incidence angle towards the lens and the center wavelength was chosen to be 225nm to match the half-reduced feature size used in the simulation (suppose 550nm is the wavelength of interest in a real case). An E-field profile and E-field movie monitor was place at the y-z plane to monitor how the light behaved after going through the lens.

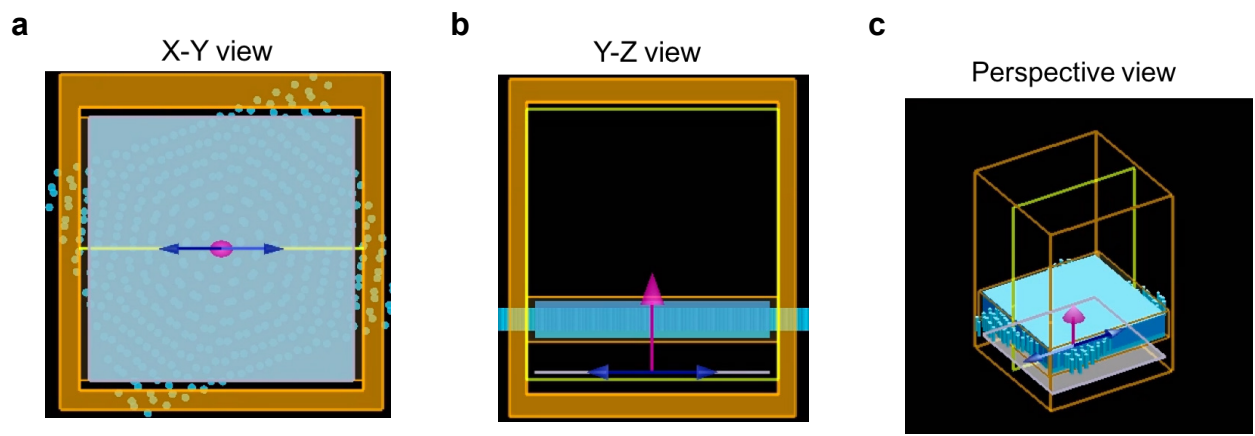

**Figure 4. Simulation model of a single micro-Mlens.** (a-c) x-y view, y-z view and perspective view respectively.

Fig. 5a shows the refractive index profile of the simulation model at the y-z plane. It is clearly seen that the occupation density of the high-refractive-index material gradually decreases from the center to the edge. The simulation results are shown in Fig. 5b-5d and demonstrated the convex lens function of the micro-Mlens. From the E-field profile shown in Fig. 5b clearly demonstrated the plane wave was focused after going through the micro-Mlens and the focal point is around 1um away from the surface of lens. Fig. 5c and 5d are two frames captured by the E-field movie monitor at y-z plane. Fig. 5c showed the E-field profile at the time when a wave packet was going towards the focal point and Fig. 5d showed the E-field profile at the time when a wave packet just passed through the focal point.

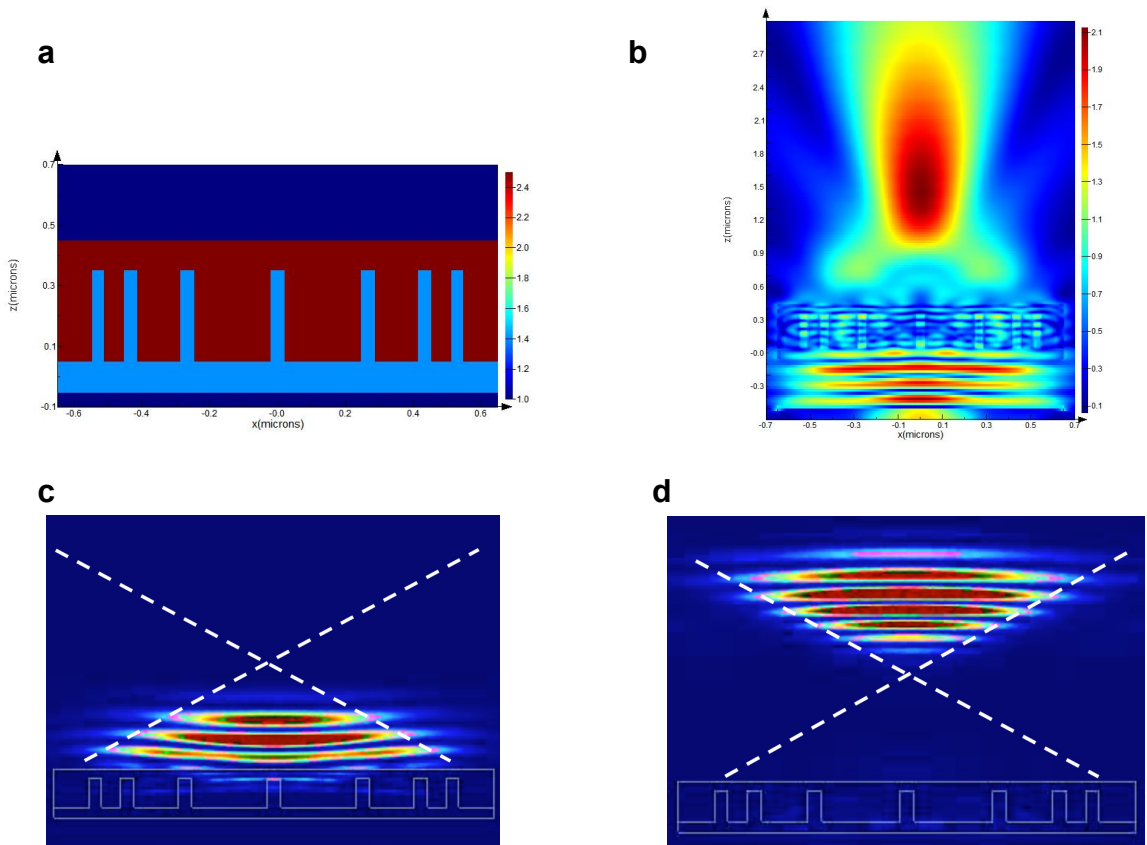

**Figure 5. Simulation results of a single micro-Mlens.** (a) refractive index profile at the y-z plane of the model. (b) E-field profile at y-z plane when a plane wave went through the micro-Mlens. (c-d) The E-field profiles captured when a plane wave packet was going towards the focal point and just passed through the focal point respectively.

**Fabrication of Mlens-array.** The Mlens-array was fabricated by double-cycle compositional nanoimprint lithography (NIL) on a 4-inch glass wafer using a hexagonal pillar array mold. And the hexagonal pillar array mold was also fabricated by NIL starting from a 1-D grating mold on a 4-inch 120nm/500um thick $SiO_2$/Si substrate and the fabrication process is schematically illustrated in Fig. 6.

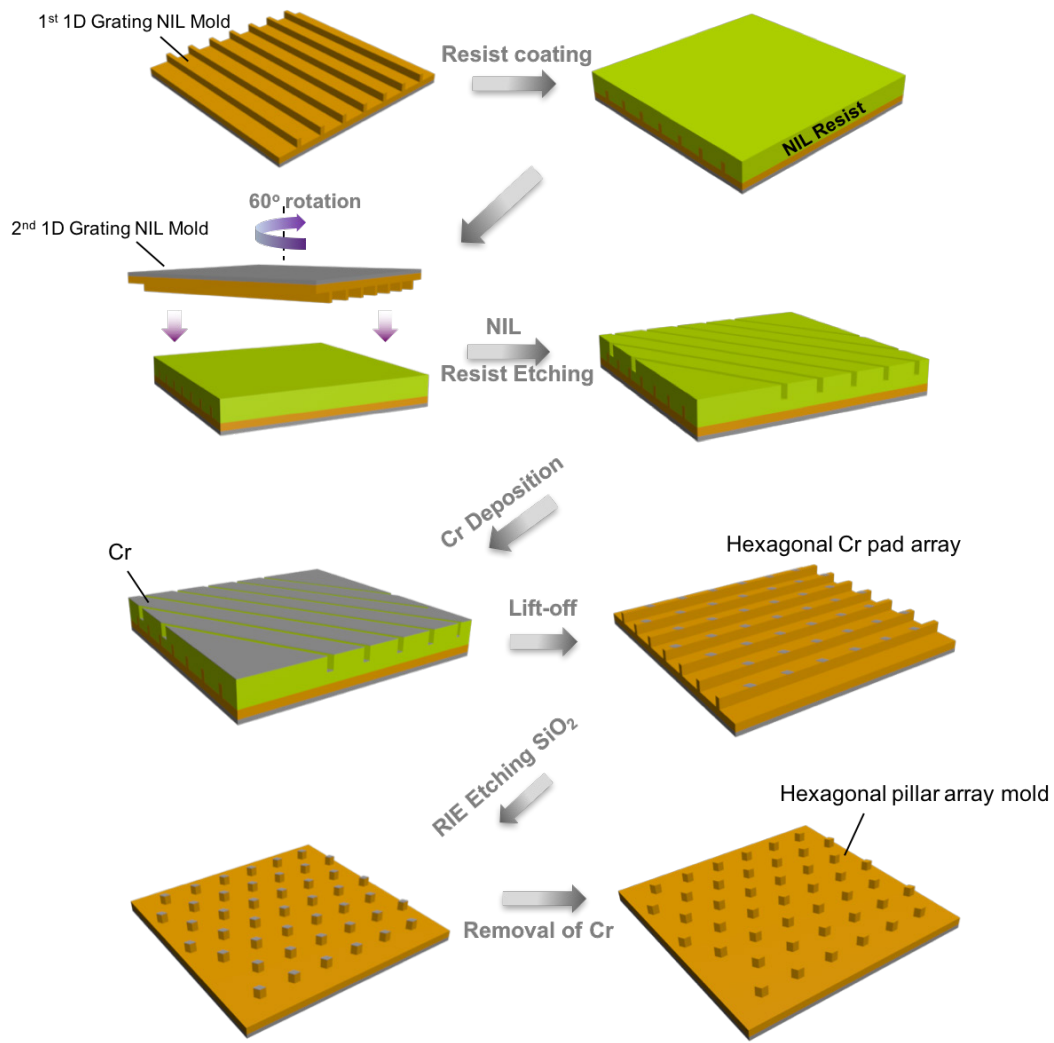

**Figure 6. Fabrication of hexagonal pillar array mold.** fabrication process: form a 1-D grating in the resist layer on top of another 1-D grating mold with relative rotation angle of 60 degrees by hot embossing. After Cr deposition, lift-off and RIE etching, a hexagonal pillar array was formed followed by Cr removal.

Firstly, 200nm thick NIL resist was spun on top of a 400nm-pitch 1-D grating mold. Then another 400nm-pitch grating mold was rotated by 60 degrees relative to the first grating mold and the grating pattern was stamped into the resist layer. After etching the residual resist layer to expose the top surface of the underneath grating mold, 10nm thick chromium (Cr) was deposited on top. Then after the lift-off process, a hexagonal array of rhombus Cr pad pattern was formed on the grids of the 1-D grating. With the Cr pattern as an etching mask, the hexagonal pillar array was finally formed by etching off the 120nm-thick unprotected $SiO_2$ using Reactive-ion etching (RIE) followed by removal the Cr mask. Additionally, the pillar size was further shrunk by wet etching in buffer HF and the mold went through further surface treatment process to make it useable.

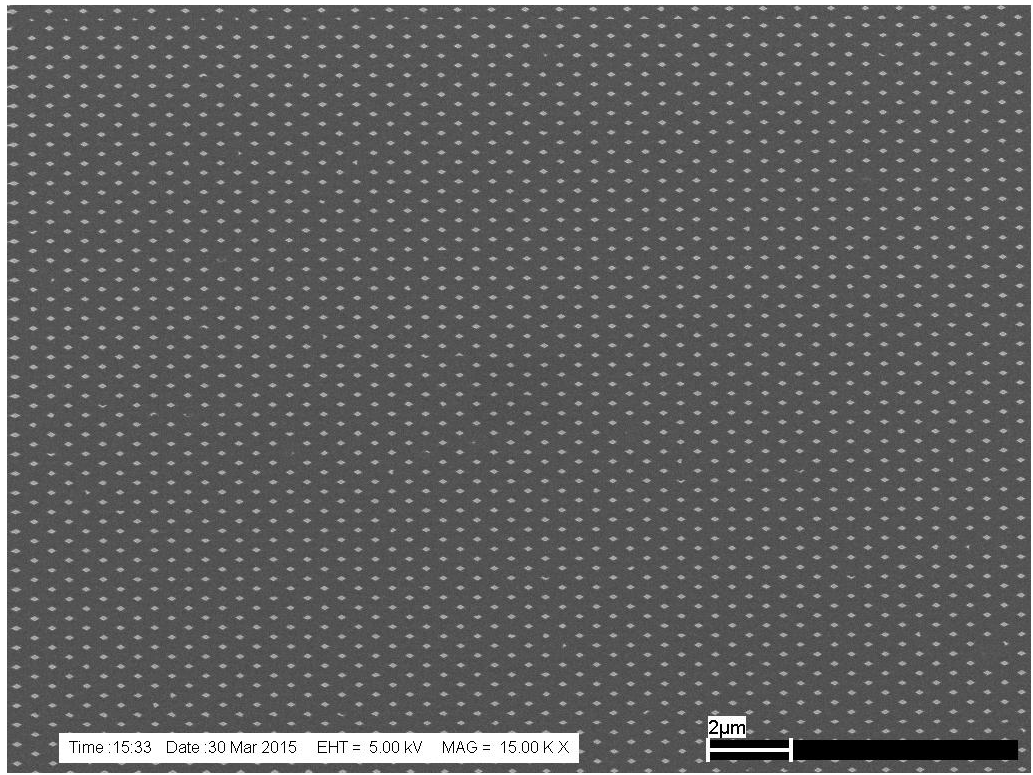

**Figure 7. Scanning electron microscopy (SEM) image showing the top view of the hexagonal pillar array mold**.

Fig.7 is the scanning electron microscopy (SEM) image of the top view of the fabricated hexagonal pillar array mold demonstrating the excellent pattern uniformity in large scale. The SEM image shows that the lattice constant of the hexagonal pattern is 461nm and the edge size of each rhombus pillar is around 70nm. Once the hexagonal pillar mold was ready, the Mlens-array was fabricated by the double-cycle compositional NIL as illustrated in Fig. 8.

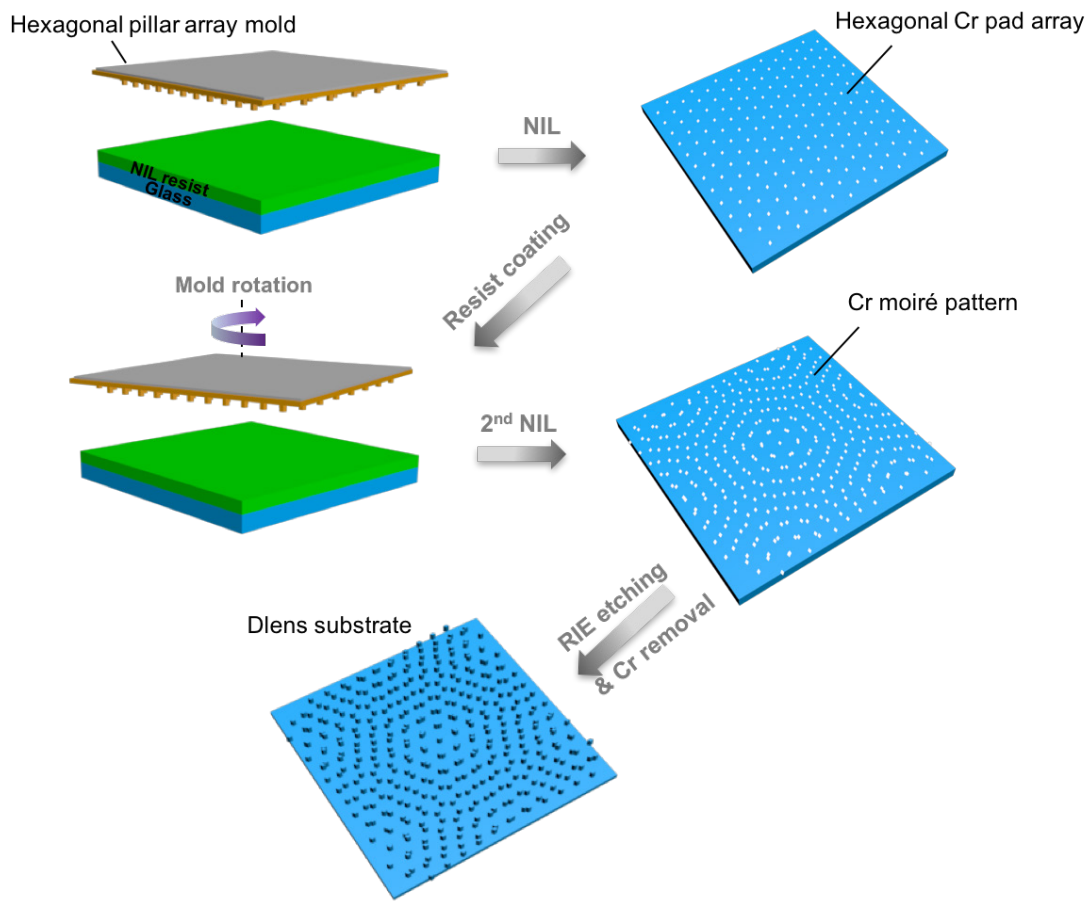

**Figure 8. Fabrication of Mlens-array.** fabrication process: a hexagonal Cr pad array was firstly patterned on the glass substrates after NIL, resist etching, Cr deposition and lift-off. Then the second hexagonal pattern was patterned overlaying on top of the first pattern with a relative

rotation angle by another NIL process. Finally, after RIE etching with Cr as mask, the Mlens-array was formed followed by removal of Cr.

A hexagonal Cr pad array was first fabricated on a 4-inch 0.5mm thick glass substrate by NIL and Cr deposition and lift-off using the just fabricated hexagonal pillar array mold. Then the second identical hexagonal Cr pad array was fabricated overlaying on top of the first Cr pad array with a relative rotation angle of 2.4 degrees by repeating the same NIL and Cr deposition and lift-off process using the same mold. Finally, with the Cr moiré pattern as an etching mask, the Mlens-array was finally fabricated by etching off 120nm thick unprotected glass using RIE followed by removal of the Cr mask. The fabricated Mlens-array was examined by SEM and the top view image is shown in Fig. 9. The period of the Mlens-array was measured to be 11um from the SEM image which is in accord with the theoretical prediction.

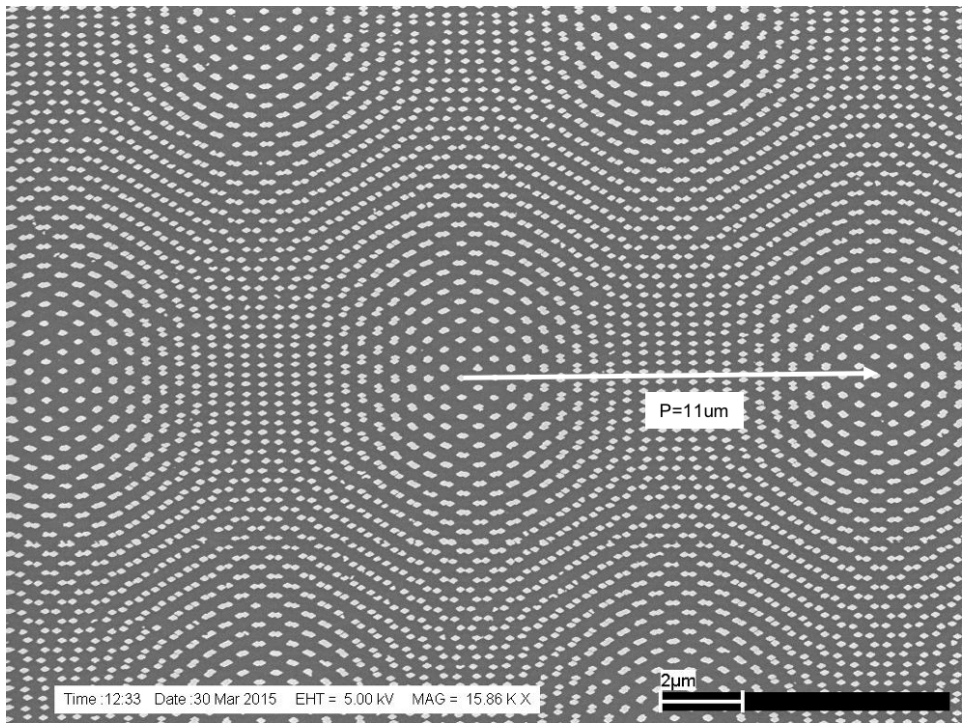

**Figure 9. Scanning electron microscopy (SEM) image showing the top view of the Mlens-array fabricated on glass substrate.**

**Over 100% light extraction enhancement of OLEDs by Mlens-array.**

**Fabrication of Mlens-array OLEDs.** One of the immediate applications of Mlens-array is for enhancing the light extraction efficiency of OLEDs. After replacing the planar glass substrate with a micro-lens array, the refraction behavior of the light rays at the glass-air interface is changed compared to a planar interface. The light rays at an incident angle larger than the critical angle of a planar glass-air interface can be coupled out by the micro-lens array. To demonstrate the light extraction enhancement of OLED by Mlens-array, we fabricated a green-emitting Mlens-array OLED by replacing the planar glass substrate with the fabricated Mlens-array substrate. The fabrication process is schematically illustrated in Fig. 10.

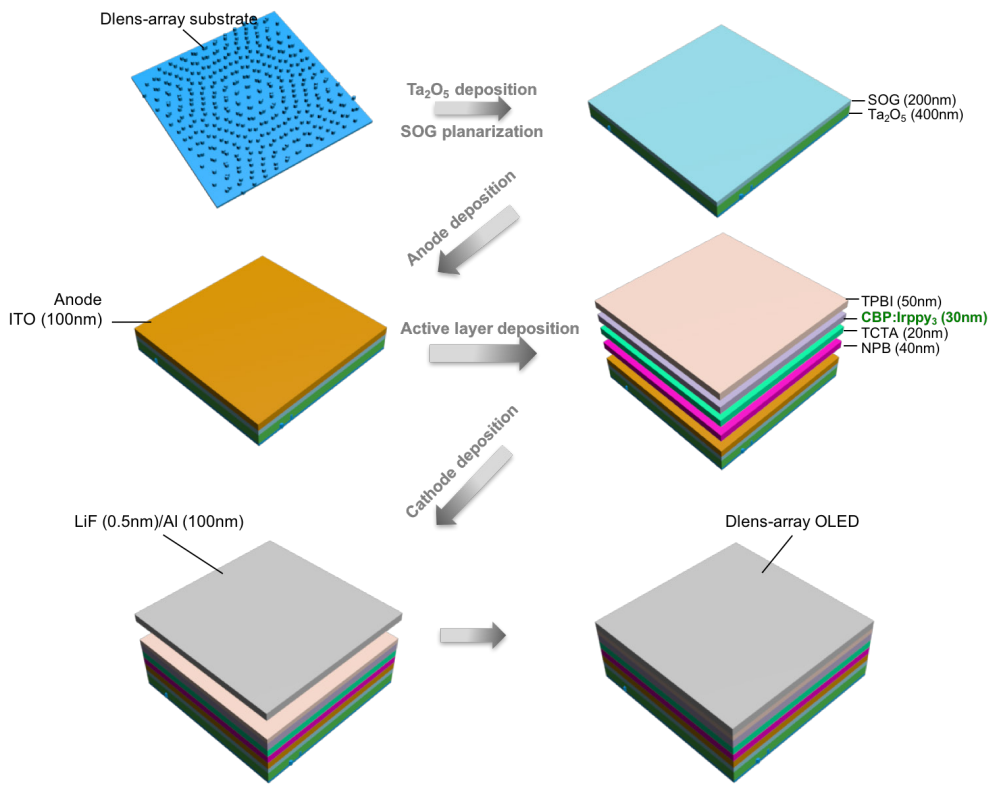

**Figure 10. Fabrication of the green-emitting Mlens-array OLED.** fabrication process: $Ta_2O_5$ was deposited on top of the glass Mlens-array substrate to form a convex micro-lens array. After the surface of $Ta_2O_5$ was planarized by a spin-on-glass layer, an ITO anode layer, organic active layers and LiF/Al cathode were sequentially deposited on top to form the Mlens-array OLED. Firstly, a 400nm-thick layer of high refractive index material $Ta_2O_5$ was deposited on top of the Mlens-array on glass substrate to form a convex micro-lens array. Then 200nm spin-on-glass was used to planarize the corrugated surface of the $Ta_2O_5$ layer. In the final step, the green-emitting Mlens-array OLED was fabricated by sequentially depositing 100nm-thick ITO layer by e-beam evaporation, 40nm-thick NPB, 20nm-thick TCTA, 30nm-thick CBP doped with 6wt% $Ir(ppy)_3$, 50nm-thick TPBI and 0.5nm/100nm thick LiF/Al by thermal evaporation under high vacuum ($<10^{-7}$ Torr). The light emitting area of a HDNM-OLED is 3mm by 3mm, which is defined by a shadow mask during the evaporation of Al back electrode. The cross-sectional SEM image (Fig. 11) shows the layer structure of the fabricated Mlens-array OLED. For comparison, reference OLEDs, "ITO-OLEDs" were also fabricated, which have the same organic active layer structure as the Mlens-array OLEDs except replacing the Mlens-array substrates with planar glass substrates.

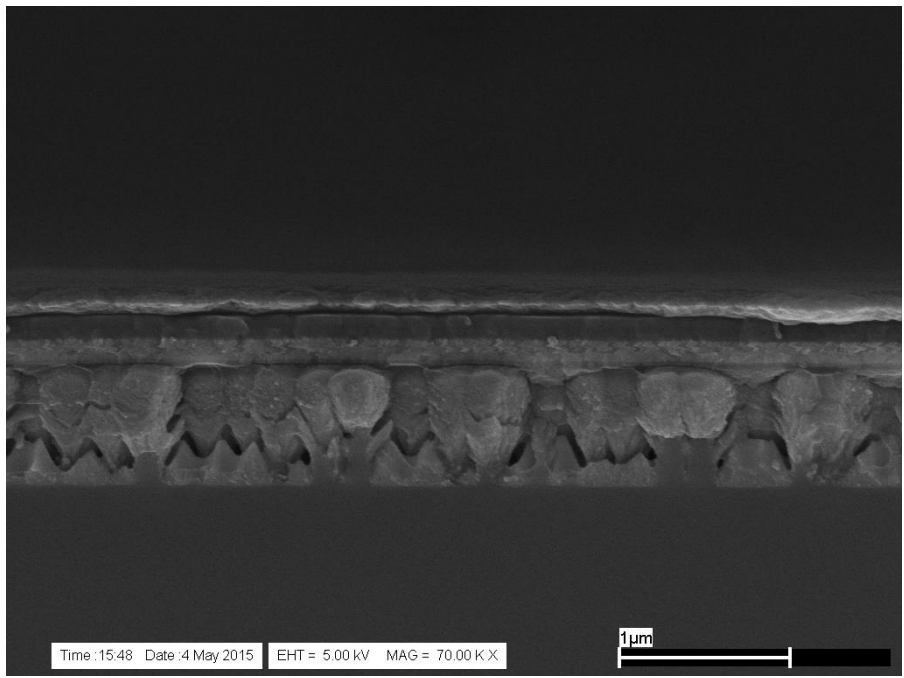

**Figure 11. Cross-sectional SEM image of the green emitting Mlens-array OLED.**

**Electroluminescence, EQE and light extraction enhancement.** The total front surface electroluminescence (EL) spectra of the green-emitting Mlens-array OLED and ITO-OLED (Fig. 12a) were measured using an integrated sphere (Labsphere LMS-100) connected to a spectrometer (Horiba Jobin Yvon). During the measurement, the devices were attached on a stage holder in the integrated sphere and the side walls of the devices were fully covered by black tapes to ensure only the light emitted from OLEDs' front surfaces was collected by the spectrometer. The measured EL spectra show that in the entire emitting range (450nm to 620nm), the EL intensity of Mlens-array OLED is significantly higher than that of ITO-OLED. And for the green-emitting material we use, the strongest emission intensity is achieved in the wavelength range of 520nm to 560nm. Compared with the ITO-OELD, at this emission peak, the Mlens-array OLED shows a 2.0-fold EL enhancement factor (i.e. the ratio of the spectrum intensity of Mlens-array OLED to ITO-OLED) and by averaging the EL enhancement factor over the entire emission wavelength range, the Mlens-array OLED shows a 1.97-fold average EL enhancement factor.

The EQE as a function of injection current density of the Mlens-array OLEDs and ITO-OLEDs (Fig. 12b) was obtained by firstly converting the measured EL spectra to power spectral densities $\rho(\lambda)$(W/nm) at different current densities using a calibrated standard lamp (Labsphere AUX-100) and then calculated by the following formula:

$$EQE = \frac{q}{hc}\int \rho(\lambda)\lambda d\lambda$$

where q is the electric charge, h is Planck constant, c is light speed in air and $\lambda$ is the wavelength.

The measured EQE shows that in the current density range from 1mA/cm² to 100mA/cm², the Mlens-array OLED exhibited a maximum EQE of 54% (at current density <2mA/cm²), which is 2.08-fold higher than that of the ITO-OELD (a maximum EQE of 26%). Considering $EQE = IQE \times \eta_{ex.}$ where $\eta_{ex.}$ is the light extraction efficiency, and we assume the IQE of the light emitting materials is not affected by the different OLED substrates, hence the light extraction efficiency enhancement should be the same as the EQE enhancement. Therefore, we demonstrated that by using the Mlens-array, the light extraction efficiency of the OLED was enhanced more than 100% which is quite remarkable.

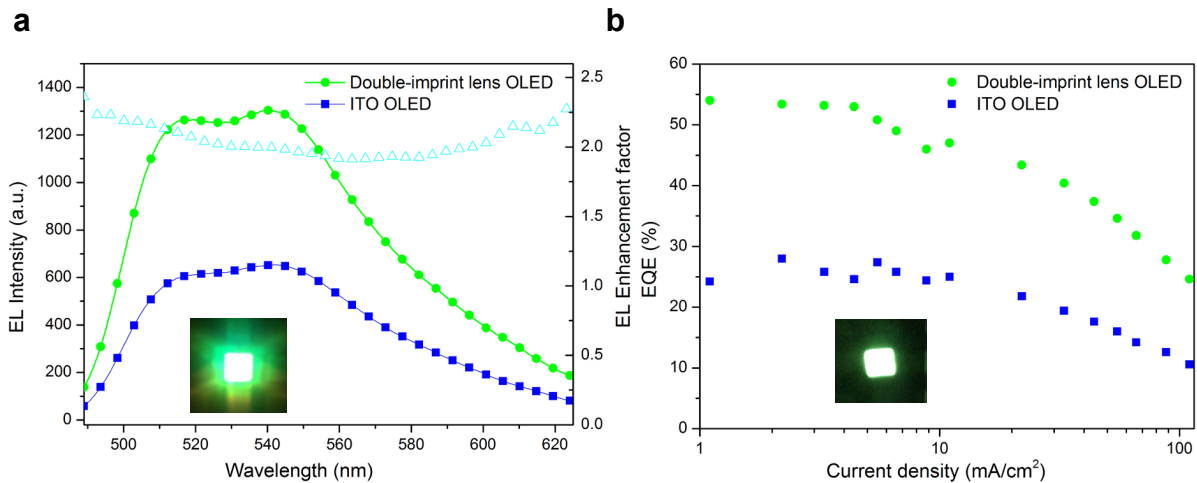

**Figure 12. Measured electroluminescence (EL) and EQE.** (a) Total front-surface EL & EL enhancement spectrum(b) EQE vs. current density. Compared with ITO-OLEDs, Mlens-array OELDs show 1.97-fold average EL enhancement factor and 2.08-fold maximum EQE enhancement factor. Insets are the optical images of the Mlens-array OLED and ITO-OLED in operation.

**Conclusion**

In this chapter, we developed a metamaterial based flat micro-lens array, termed "Mlens-array", consisting of a hexagonal moiré pattern pillar array. By choosing a low refractive index dielectric material for the pillar array and a high refractive index dielectric material for the layer capped on top of the pillar array, we fabricated the Mlens-array behaving as a conventional convex optical micro-lens array. Using the FDTD simulation software, we demonstrated the focusing function of a single micro-lens of Mlens-array for plane wave and hence verified the convex-lens behavior of the designed Mlens-array. The Mlens-array was fabricated on a 4-inch wafer-size glass substrate by double-cycle compositional nanoimprint lithography which is easy for achieving high throughput fabrication in large-scale. And as one of the most promising applications, we also demonstrated that by applying the Mlens-array substrate in a typical green-emitting OLED, the light extraction efficiency was enhanced by over 100% (2.08-fold) compared to a control device fabricated on the conventional planar glass substrate.